\documentclass[%
 preprint,
superscriptaddress,
 amsmath,amssymb,
 aps,
pra,
floatfix,
]{revtex4-2}

\usepackage{graphicx}
\usepackage{dcolumn}
\usepackage{bm}
\usepackage{amsmath}
\usepackage{float}
\usepackage{multirow}

\begin{document}

\preprint{APS/123-QED}

\title{\textbf{Small-Angle Differential Cross Sections for Symmetrical
Resonant Charge Exchange in Molecular Hydrogen}}

\author{Jibak Mukherjee}
\affiliation{
Tata Institute of Fundamental Research, Mumbai 400005, India
}

\author{Kamal Kumar}
\affiliation{
Tata Institute of Fundamental Research, Mumbai 400005, India
}

\author{Harpreet Singh}
\affiliation{
Tata Institute of Fundamental Research, Mumbai 400005, India
}

\author{Manojit Das}
\affiliation{
Tata Institute of Fundamental Research, Mumbai 400005, India
}

\author{Deepankar Misra}
\email{dmisra@tifr.res.in}
\affiliation{
Tata Institute of Fundamental Research, Mumbai 400005, India
}

\date{\today}

\begin{abstract}
\end{abstract}

\begin{abstract}
We measured state selective scattering angle distributions in non-dissociative-state single electron capture in $\lbrace 5, 7.5, 10\rbrace$ keV/u, $\left[\mathrm{H^+ + H_2}\right]$ and $\left[\mathrm{H_2^+ + H_2}\right]$ collisions, using Cold Target Recoil Ion momentum Spectroscopy (COLTRIMS). In our measurements, range of the scattering angle was small where diffraction effect is expected to take place, and, the ground state electron capture - symmetrical resonant channel for the $\mathrm{H_2^+}$ projectile and near resonant channel for the $\mathrm{H^+}$ projectile - was fully resolved from other channels. We analyzed the data considering $\mathrm{H_2^+}$ and $\mathrm{H_2}$ possess two independent-identical scattering centers and exhibit multi center interference effect. By comparing these equal-velocity collisions, we investigated the atomic nature of the individual scattering centers. We observed Fraunhofer diffraction from a circular aperture like patterns with equal fringe widths, for both the projectiles, in ground state channel. We applied a simple toy model to reconstruct the circular apertures.
\end{abstract}

\maketitle
\date{\today}

\maketitle


\section{\label{sec:intro}Introduction}
In ion-atom/molecule collision, charge transfer (electron capture) is called `symmetrical resonance' when the reactant and the products remain same, before and after the collision. This can be represented as $\mathrm{A^+(\Psi) + A(\Phi) \rightarrow A(\Phi) + A^+(\Psi)}$, where $\mathrm{A^+}$ and A are the two colliding particles and $\mathrm{\Psi}$ and $\mathrm{\Phi}$ represents bound electronic states. This was a highly studied problem in previous century (see for example \cite{10.1063/1.1743499},\cite{10.1063/1.1733066,PhysRev.131.229} and references therein). Superposition of pathways between symmetrical and anti-symmetrical adiabatic states  are primarily responsible for oscillatory charge transfer probability that undulate with collision velocity at a fixed impact parameter, or with impact parameter at a fixed collision velocity. This kind reaction was first experimentally observed by Ziemba and Everhart in $\mathrm{He^+ + He}$ collision \cite{PhysRevLett.2.299}, at large scattering angle. Later in different collision systems, for example $\mathrm{H^+ + H}$\cite{PhysRev.125.567}, $\mathrm{He^+ + He}$\cite{PhysRev.132.2078}, $\mathrm{Ar^+ + Ar}$ and $\mathrm{Ne^+ + Ne}$\cite{PhysRev.147.76}, symmetrical resonance was experimentally investigated with more details. A distinct oscillatory characteristics was found in charge transfer probability at large scattering angle regime where classical treatment is possible and the collision take place at very small impact parameter. 

Measurement on molecular systems present several challenges.   Following charge transfer, a molecular ion may undergo dissociation, participate in a chemical reaction, or remain a neutral molecule. For example, in collision of $\mathrm{H_2^+}$ with $\mathrm{H_2}$, the products may include two neutral hydrogen atoms, the $\mathrm{H_3^+}$ ion or neutral $\mathrm{H_2}$\cite{10.1063/1.555858,PhysRev.130.1852}. Because of this complexity, oscillations in the symmetric resonant charge transfer cross section have not been properly observed \cite{PhysRev.118.1552}. 

Molecular ions have rotational and vibrational degrees of freedoms. In fully quantum or semiclassical treatment, calculation incorporating these degrees of freedoms add extra complexities. This problem was dealt with different modeling. In some notable works,  Gurnee and Magee\cite{10.1063/1.1743499} used Heitler-London functions to the write electronic wave function and predicted that molecular charge transfer probability as a function of impact parameter also manifest sinusoidal behavior like atomic resonant transfer case and the interaction Hamiltonian will contain additional terms coming from vibrational and rotational overlap integrals. Kimura \cite{PhysRevA.32.802} applied diatoms-in-molecule (DIM) method that was first introduced by Ellison \cite{10.1021/ja00905a003}, in $\mathrm{H^+ + H_2}$ collision system to obtain molecular wavefunction followed by coupling calculation. However, only total cross section measurements were available to testify these theories \cite{Schmid1961,10.1098/rspa.1960.0116,PhysRev.130.1852,Keene01041949}. More detailed reaction dynamics can be understood from impact parameter dependence that required differential cross section measurements. 

Our general understanding on scattering theory dictates that when deBroglie wavelength of the incident particle became much smaller than internuclear separation, two nuclei in diatomic molecule start acting like two independent - coherent scattering centers\cite{book3}. Also at keV energy range, vibrational and rotational timescales is much smaller than the transit time of the colliding particle, thus "adiabatic sudden approximation" can be applied.  As a result, Young's-double slit interference effect manifest in the scattering processes  \cite{PhysRev.150.30,doi:10.1073/pnas.1018534108}. In charge exchange reaction, existence of this interference effect was first predicted by Tuan and Gerjuoy \cite{PhysRev.117.756} in $\mathrm{H^+ + H_2}$ collision and later investigated by many works by treating the molecul and verified by many theoretical \cite{PhysRevA.23.1807,PhysRevA.40.1302,PhysRevA.40.3673,PhysRevA.38.3769} and experimental \cite{PhysRevA.47.3923,PhysRevLett.101.083201,PhysRevLett.102.153201,PhysRevLett.101.173202,PhysRevLett.112.023201} studies. The electronic wavefunction of a homonuclear diatomic molecule, under LCAO framework can be written as $\mathrm{\Psi_m = \Psi_a(1) \pm \Psi_a(2)}$, where $\mathrm{\Psi_a}$ represents atomic wavefunction centered around two nuclei, $\pm$ sign stands for gerade and ungerade states. $\mathrm{\Psi_a}$ can be expanded in superposition of stationary states of constituent atom. In $\mathrm{H_2^+ + H_2}$ resonant charge transfer, electronic states remains same but rotational and vibrational levels may be different, before and after collision. If we can characterize scattering properties of $\Psi_a$ in $\mathrm{H_2^+}$  and $\mathrm{H_2}$, we can convert the molecular scattering into effective atomic scattering problem and get comprehensive information about the reaction dynamics, assuming vibrational and rotational levels have uniform effect over the whole impact parameter range. 

In our experiment we measured state selective differential cross sections in $\mathrm{H_2^+ + H_2}$ collisions. We used state of the art Cold Target Recoil Ion Momentum Spectroscopy (COLTRIMS) \cite{DORNER200095} momentum imaging techniques to unambiguously measure the recoil momentum vector embedding scattering angle and Q-value information, for resonant capture channel. According to impact parameter model, in intermediate velocity range, relative collision velocity governs the time dependent perturbation to the electronic Hamiltonian \cite{book1,book2}. At the nascent stage in modeling the identical scattering centers in $\mathrm{H_2^+}$, we compared the state selective differential cross sections with those of equal velocity $\mathrm{H^+}$ projectile.  In small angle scattering, eikonal approximation is highly applicable \cite{PhysRev.169.84,R_McCarroll_1968}. This method describes the scattering processes in Fraunhofer diffraction analogy \cite{M_van_der_Poel_2002,PhysRevLett.87.123201}. In this analogy, atomic scattering can be modeled with as a Fraunhofer type diffraction from a circular aperture where impact parameter dependence is imprinted. The identical scattering centers in $\mathrm{H_2^+}$ can also be modeled with an identical circular aperture which will explain the resonant charge transfer at a particular collision velocity. By comparing equal velocity $\mathrm{H_2^+}$ and $\mathrm{H^+}$ projectiles, we are making a comparison between two circular apertures whom represents an isolated atomic ion and an effective atomic ion in molecular environment. 
\section{\label{sec:theory}Theory}
\subsection{Multi center interference effect}
In our experiment, projectile velocities were 0.447, 0.548, 0.633 a.u.. At these velocities, de-Broglie wavelength is  $\mathrm{\lambda_{dB}\sim10^{-13}}$ m, and the interaction timescale is of the order $\mathrm{10^{-16}}$ sec.. Thus,  in $\mathrm{H_2}$ rest frame, the incident projectile matterwave see the target as frozen two scattering center object and manifest two center interference effect \cite{PhysRev.117.756,PhysRevLett.101.083201,PhysRevLett.102.153201}.  In $\mathrm{H_2^+}$ rest frame, the incident target matterwave also see the projectile as a molecular double slit at a fixed orientation and exhibit Young double slit interference pattern \cite{PhysRevLett.101.173202,PhysRevLett.112.023201}. A corollary can be drawn that, in our experiment, $\mathrm{H^+ + H_2}$ collision will exhibit two center interference and $\mathrm{H_2^+ + H_2}$ collision will exhibit effective four center interference (two in projectile and two in target) effects. In Born series, first order T-matrix can be written as \cite{book3,PhysRev.117.756, PhysRevA.38.3769,PhysRevA.23.1807,PhysRevA.40.1302,PhysRevA.40.3673}
\begin{align}
   T^{H^+}_{H_2} &= T_0(1+e^{i\Delta \boldsymbol{K}\cdot\boldsymbol{R_{H_2}}})
    \label{eqn:TH+H2} \\
   T^{H_2^+}_{H_2} &= T_{00}(1+e^{i\Delta \boldsymbol{K}\cdot\boldsymbol{R_{H_2^+}}})(1+e^{i\Delta \boldsymbol{K}\cdot\boldsymbol{R_{H_2}}}) \label{eqn:TH2++H2} 
\end{align}
    
where $\mathrm{T^{H^+}_{H_2}}$ and $\mathrm{T^{H_2^+}_{H_2}}$ are the collision system T-matrix, $\mathrm{T_0}$ and $\mathrm{T_{00}}$ represents single center T-matrix, associated with  projectile-target one scattering center pair, in $\mathrm{H^+ + H_2}$ and $\mathrm{H_2^+ + H_2}$ collision systems, respectively.   ${\boldsymbol{\Delta K = \Delta K_\parallel + \Delta K_\perp}}$ is the change in wave vector in the scattering, $\mathrm{\boldsymbol{R_{H_2}}}$ and $\mathrm{\boldsymbol{R_{H_2^+}}}$ are the internuclear separation of $\mathrm{H_2}$ and $\mathrm{H_2^+}$ during the collision. 
    
For non-dissociative electron capture, the measured  differential cross sections will be molecular orientation averaged and the differential cross section following Eqn. \ref{eqn:TH+H2}, \ref{eqn:TH2++H2} can be written as \cite{book3}
\begin{align}
    \left<\dfrac{d\sigma}{d\Omega}\right>^{H^+}_{H_2} &= \left<\dfrac{d\sigma_0}{d\Omega}\right>I(\Omega,R_{H_2})\label{eqn:dndo_H+}, \\
\left<\dfrac{d\sigma}{d\Omega}\right>^{H_2^+}_{H_2} &= \left<\dfrac{d\sigma_{00}}{d\Omega}\right>I(\Omega,R_{H_2^+}, R_{H_2})\label{eqn:dndo_H2+},\\
I(\Omega,R_{H_2^+}, R_{H_2}) &= I(\Omega,R_{H_2^+})I(\Omega,R_{H_2}) \label{eqn:intf_H2+},\\
I(\Omega,R_{H_2})&=\left( 1+\dfrac{sin(\Delta K R_{H_2})}{\Delta K R_{H_2})}\right)\label{eqn:intf_H2},\\
I(\Omega,R_{H_2^+})&=\left( 1+\dfrac{sin(\Delta K R_{H_2^+})}{\Delta K R_{H_2^+})}\right),
\end{align}
where $\mathrm{\left<\dfrac{d\sigma}{d\Omega}\right>^{H^+}_{H_2}}$ and $\mathrm{\left<\dfrac{d\sigma}{d\Omega}\right>^{H_2^+}_{H_2}}$ are the orientation averaged (integrated over all $\boldsymbol{\Delta K\cdot R}$) differential cross sections for $\mathrm{H^+}$ and $\mathrm{H_2^+}$ projectiles, colliding with $\mathrm{H_2}$ target. Here we define the terms $\mathrm{\left<\dfrac{d\sigma_0}{d\Omega}\right>}$ and $\mathrm{\left<\dfrac{d\sigma_{00}}{d\Omega}\right>}$ as the single center diffraction terms. The interference effect will be manifested by$\mathrm{I(\Omega, R_{H_2})}$ and $\mathrm{I(\Omega, R_{H_2^+}, R_{H_2})}$ terms, hence we define them as the interference terms in $\mathrm{H^+ + H_2}$ and $\mathrm{H_2^+ + H_2}$ collision systems. 

\subsection{Application of the Impact parameter model with eikonal approximation}
In this semiclassical description, nuclear trajectories are given as 
\begin{equation}
    \boldsymbol{R = b+ v_0}t;\  \boldsymbol{v_0\cdot b=0}
\end{equation}
where b is the impact parameter, $\mathrm{v_0}$ is relative collision velocity and with these rectilinear trajectories 
electronic motion is obtained by solving Schr\"odinger equation\cite{book2,book1,PhysRev.169.84,R_McCarroll_1968,10.1098/rspa.1966.0126}. The differential cross section is given by
\begin{align}
f_{i\rightarrow f}(\boldsymbol{K_\perp}) &= -i\dfrac{\mu v_0}{2\pi} \int d^2\boldsymbol{b}e^{i\boldsymbol{K_\perp}\cdot\boldsymbol{b}}\tilde{\mathcal{A}}_{i\rightarrow f}(\boldsymbol{b})e^{2i\delta(\boldsymbol{b})},\label{eqn:imp 2d}\\
    \dfrac{d\sigma_{i\rightarrow f}}{d\Omega}(\boldsymbol{K_\perp}) &= |f_{i\rightarrow f}(\boldsymbol{K_\perp})|^2,\label{eqn:dndo}
\end{align}
where $\mathrm{\mu}$ is the reduced mass of the target and projectile system, $\mathrm{\mathcal{\tilde{A}}}_{i\rightarrow f}(b)$ is the probability amplitude, $\mathrm{e^{2i\delta_{i\rightarrow f}(b)}}$ is the eikonal phase.
For atomic projectile-target collision, the interaction is 
cylindrically symmetric. In this case the scattering amplitude is 
\begin{equation}
   f_{i\rightarrow f}(\boldsymbol{K_\perp}) = -i\mu v_0 e^{-i\Delta m \Phi} (-i)^{\Delta m}\int_0^\infty\mathcal{\tilde{A}}_{i\rightarrow f}(b)e^{2i\delta_{i\rightarrow f} (b)}J_{\Delta m}(\mu v_0\Theta b)bdb, \label{eqn:fth} 
\end{equation}

where  $\mathrm{\Delta m = m_f - m_i}$ is the change in magnetic quantum number of the reaction, $\mathrm{J_{\Delta m}}$ is the Bessel function of first kind. In a channel, if $\mathrm{\Delta m =0}$, the simplified expression is given by
\begin{equation}
   f_{i\rightarrow f}^{\Delta m =0}(\boldsymbol{K_\perp}) = -i\mu v_0 \int_0^\infty\mathcal{\tilde{A}}_{i\rightarrow f}(b)e^{2i\delta_{i\rightarrow f} (b)}J_{0}(\mu v_0\Theta b)bdb, \label{eqn:fthdm0} 
\end{equation}

\subsubsection{A toy model for channel with $\mathrm{\Delta m =0}$}
At high energy, $small\ angle$ scattering, Eqn. \ref{eqn:imp 2d} works very well. For atomic collision, Eqn, \ref{eqn:fthdm0} has an optical analogy, the inelastic scattering is equivalent to matter wave diffraction from a circular aperture, characterized by $\tilde{\mathcal{A}}(\boldsymbol{b})e^{2i\delta (\boldsymbol{b})}$ \cite{PhysRevLett.87.123201,M_van_der_Poel_2002}. If the individual scattering centers in molecular collision acts like identical atomic scatterers then we can replace each scattering centers with a circular aperture and construct the double slit for $\mathrm{H^+ + H_2}$ and quadrouple slit for $\mathrm{H_2^+ + H_2}$. Mathematically, we can model $\mathrm{T_0}$ and $\mathrm{T_{00}}$ of Eqn. \ref{eqn:TH+H2} and \ref{eqn:TH2++H2} with Eqn. \ref{eqn:fthdm0}. In orientation averaged measurement, we observe the diffraction pattern of the atomic circular aperture through the terms  $\mathrm{\left<\dfrac{d\sigma_0}{d\Omega}\right>}$ and $\mathrm{\left<\dfrac{d\sigma_{00}}{d\Omega}\right>}$ in Eqn. \ref{eqn:dndo_H+} and \ref{eqn:dndo_H2+}. The circular aperture satisfy the equation
\begin{equation}
    \left<\dfrac{d\sigma_{0,00}}{d\Omega}\right> = n_c \left|\int_0^{b_{max}} \mathfrak{\tilde{A}}_{i\rightarrow f}(b)e^{2i\delta_{i\rightarrow f} (b)}J_0(\mu v_0 \Theta b) b db\right|^2, \label{eqn:tm1}
\end{equation}
    where $\mathcal{\tilde{A}}(b)e^{2i\delta (b)}$ is represented as $\mathfrak{\tilde{A}}(b)$, $\mathrm{n_c}$ is a normalizing constant, $\mathrm{b_{max}}$ is a limiting value beyond which capture does not take place. Obtaining the circular aperture, the double and quadruple slits will be written as \cite{PhysRevA.40.3673,PhysRevA.40.1302}
\begin{align}
    \mathfrak{\tilde{A}}_{H_2}^{H^+}(\boldsymbol{B}) &= \mathfrak{\tilde{A}}(\boldsymbol{b_1}) + \mathfrak{\tilde{A}}(\boldsymbol{b_2})e^{i\Delta K_\parallel R_{H_2}^{\parallel}}, \label{eqn:ab2d h + h2} \\
    \mathfrak{\tilde{A}}_{H_2}^{H_2^+}(\boldsymbol{B}) &=  \mathfrak{\tilde{A}}(\boldsymbol{b_1}) +  \mathfrak{\tilde{A}}(\boldsymbol{b_2})e^{i\Delta K_\parallel R_{H_2}^\parallel} +  \mathfrak{\tilde{A}}(\boldsymbol{b_3})e^{i\Delta K_\parallel R_{H_2^+}^\parallel} +  \mathfrak{\tilde{A}}(\boldsymbol{b_4})e^{i\Delta K_\parallel R_{H_2}^\parallel + i K_\parallel R_{H_2^+}^\parallel} \label{eqn:ab2d h2+ + h2},
\end{align}
where $\boldsymbol{B}$ is the impact parameter of $\mathrm{H_2^+}$ center of mass, with respect to $\mathrm{H_2}$ center of mass, $\lbrace\boldsymbol{b_i}; i=1,4\rbrace$, represents impact parameters of individual projectile nucleus with respect to individual target nucleus.

 Our model numerically derive the circular aperture by inverse curve fitting method. The term inside square moduli of Eqn. \ref{eqn:tm1} is a Hankel transformation equation, hence in principle it is possible to obtain the quantity $\tilde{\mathfrak{A}}(b)$ by performing an inverse Hankel transformation. To this end, we first expanded the quantity in Fourier Bessel series
\begin{equation}
            \mathfrak{\tilde{A} (b)} = \sum_{k=1}^{n} (\alpha_k + i \beta_k)J_0\left(\dfrac{z_k b}{b_{max}}\right),
            \label{eqn:tm2}
\end{equation}
 
where $\mathrm{\alpha_k}$ and $\mathrm{\beta_k}$ are independent, real valued parameters, $\mathrm{z_k}$ is $k^{th}$ zero of Bessel function $\mathrm{J_0}$. By setting the initial guess 
\begin{align}
    |\mathfrak{\tilde{A}}(b)| &= \int_0^\infty \sqrt{\left.\dfrac{d\sigma}{d\Omega}\right|}_{meaasured},\\
    arg(\mathfrak{\tilde{A}}(b)) &= \dfrac{2log(b)}{v_0},
\end{align}
we performed least square fitting. From the optimized parameters $\lbrace \alpha_k, \beta_k\rbrace$, we constructed $\mathfrak{\tilde{A}}(b)$ and from that, the double and quadrupole slits following Eqn. \ref{eqn:ab2d h + h2} and Eqn. \ref{eqn:ab2d h2+ + h2}.

 \begin{figure}[h]
     \centering
     \includegraphics[width=\linewidth]{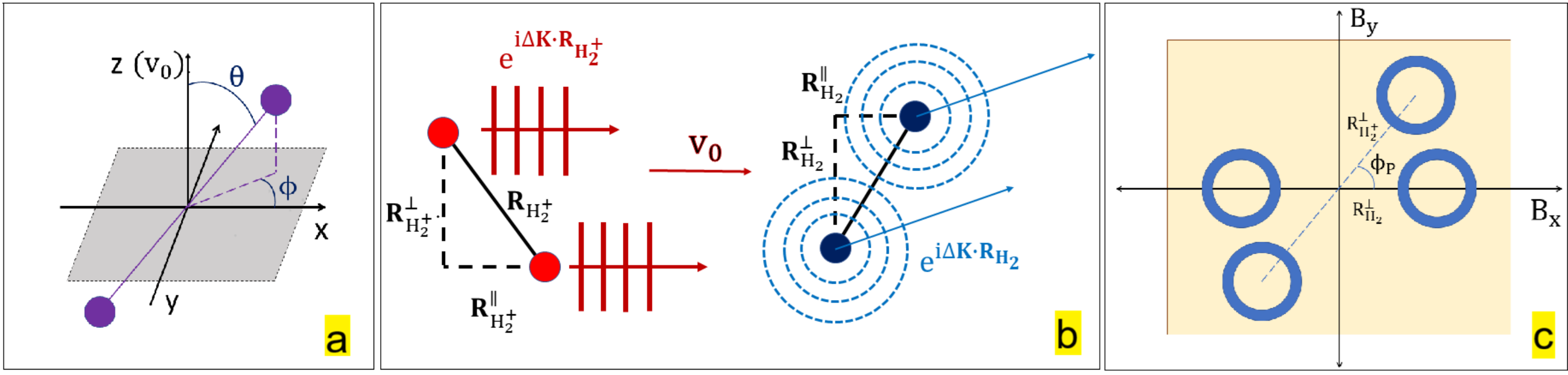}
     \caption{Illustration of matterwave Fraunhofer diffraction. Panel (a) : molecular orientation defined in polar coordinate system taking Z axis parallel to the relative velocity vector $\mathrm{v_0}$. Panel (b) : Schematic of $\mathrm{H_2^+ + H_2}$ scattering. In $\mathrm{H_2}$ rest frame, the incident $\mathrm{H_2^+}$ matterwave  will see the two nuclei of $\mathrm{H_2}$ as independent scattering centers. Also the two nuclei of $\mathrm{H_2^+}$ will act as independent wave packets with plane wavefront. Panel (c) : Illustration of Eqn. \ref{eqn:ab2d h2+ + h2}. $\mathrm{H_2}$ is in  x-z plane, $\mathrm{H_2^+}$ is in a rotated x-z plane along z axis, by an angle $\mathrm{\phi_P}$. The quadruple slit is constructed by placing identical circular apertures, represented with blue rings. Projection of internuclear separations on the impact parameter plane set the distances among them.}
     \label{fig:diffraction}
 \end{figure}


 
\section{\label{sec:exp}Experimental Setup}
To perform the experiment, we obtained ion  beam from ECRIA facility at TIFR, Mumbai \cite{Agnihotri_2011,10.1063/1.4738642}. Both the $\mathrm{H^+}$ and $\mathrm{H_2^+}$ ions were created in a 14.5 GHz Electron Cyclotron Resonance (ECR) ion source from $\mathrm{H_2}$ gas. They were extracted with desired voltages which gave intended projectile energy. Then they are selected for the mass over charge (m/q) using a $\mathrm{90^0}$ dipole magnet. Before entering the collision chamber, the projectile ion beam was further cleaned by a pair of $\mathrm{30^0}$ electrostatic cylindrical electrodes. Inside the collision chamber the projectile ion beam collided with the cold target which was prepared using supersonic expansion of $\mathrm{H_2}$ gas through a 30 $\mathrm{\mu m}$ nozzle and was passed through two skimmers for geometric cooling purpose. In the collision region, estimated parallel temperature of the supersonic jet was $\mathrm{T_\parallel}\sim$ 6K. After the collision, the non-interacting projectile was separated from the electron captured projectile using an electrostatic charge state deflector. After single electron capture, the neutral projectile  hit a time and position sensitive projectile detector (MCP + DLD from Roentdek), mounted approximately 1m downstream. The change in projectile momentum  was measured using momentum spectrometer \cite{10.1063/1.4916680,10.1063/5.0100395}. For this purpose, the recoil ion was extracted using a 5.33 V/cm electric field. After traversing an acceleration region and a field free drift region, the recoil ion hit a time and position sensitive ion detector (MCP + HEX from Roentdek). 

The estimated recoil ion momentum resolution was $\mathrm{\delta p_x}\sim$ 0.2 a.u., $\mathrm{\delta p_y}\sim$ 0.6 a.u., $\mathrm{\delta p_z}\sim$ 0.2 a.u., where the projectile beam moved along x-axis, supersonic jet traveled along y-axis and the recoil ion was extracted along z-axis. Raw data from the recoil ion and the projectile detector were analyzed in coincidence. From the measured recoil ion momentum vector components, the observables were calculated using the relations
\begin{align}
    Q &= -p_\parallel v_0 - v_0^2/2 \\
    \theta &= \dfrac{p_\perp}{m_p v_0}
\end{align}
 where $\mathrm{p_\parallel}$ is the x-momentum component ($\mathrm{p_x}$),  Q is the Q-value in the inelastic collision (total binding energy difference between initial and the final collision system), $\mathrm{\theta}$ is the scattering angle in the lab frame, $\mathrm{v_0}$ is the relative velocity between the projectile and the target, $\mathrm{m_p}$ is the mass of the projectile ion. The transverse momentum ($\mathrm{p_\perp}$) distribution was derived by inverse Abel transformation \cite{Alessi_2015}. 
 
\section{Results and discussion}
\subsection{Q-value spectrum}
\begin{figure}[h]
    \centering
    \includegraphics[width=\linewidth]{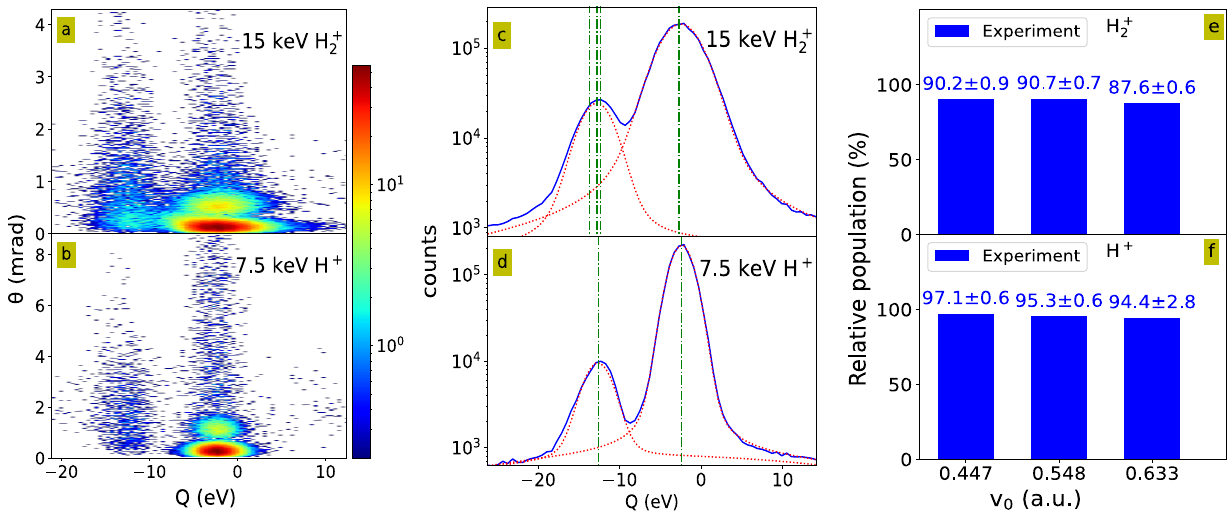}
    \caption{Panel (a, b) : measured differential spectrum for 7.5 keV/u projectiles. Panel (c,d) : measured Q-value spectrum. Contribution of two resolved peaks are shown in magenta dotted curve and calculated Q-values for electron capture to the electronic states, listed in table \ref{tab:q} are shown in green dashed vertical lines. Panel (e, f): relative population of ground state following charge transfer.}
    \label{fig:Q}
\end{figure}
We observed two resolved channels in both  $\mathrm{H^+ + H_2}$ and $\mathrm{H_2^+ + H_2}$ collisions. For $\mathrm{H_2^+}$ projectile the channels are only for non-dissociative state electron capture. The measured fully differential spectrum for 15 keV $\mathrm{H_2^+}$ and 7.5 keV $\mathrm{H^+}$ projectiles are shown in fig. (\ref{fig:Q} (a)) and fig. (\ref{fig:Q} (b)), respectively. The Q-value spectrum for the same projectiles are plotted in fig. (\ref{fig:Q} (c)) and fig. (\ref{fig:Q} (d)). Q-value is defined as total binding energy difference between the initial and the final collision system. Cold $\mathrm{H_2}$ target was always in ground state ($\mathrm{H_2(X^1\Sigma_g^+; \nu = 0)}$. Since inside the ECR ion source, $\mathrm{H_2^+}$ was created by electron impact ionization, we assumed before collision $\mathrm{H_2^+}$ was dominantly in the state $\mathrm{H_2^+(X^2\Sigma_g^+; \nu = 2)}$. To identify the electronic state populated following electron capture, we used theoretical energy level data reported by Fantz and W\"underlich \cite{FANTZ2006853}. We also assumed Frank-Condon principle was being obeyed and to calculate Q-value we took binding energy of the most probable vibrational level  of different electronic states, given by Frank-Condon factors\cite{WUNDERLICH2011152}, as a representative final state binding energy. The probable populated states following electron capture are listed in table \ref{tab:q}. Also their representative Q-values are marked with green dash-dot vertical lines in fig. (\ref{fig:Q}(c) and \ref{fig:Q}(d)). The most dominating non-dissociative state after electron capture was $\mathrm{H_2(X^1\Sigma_g^+)}$ in $\mathrm{H_2^+ + H_2}$ collision and H(n=1) in $\mathrm{H^+ + H_2}$ collision. The other states were $\mathrm{H_2 : (EF^1\Sigma_g^+, C^1\Pi_u, c^3\Pi_u, a^3\Sigma_g^+, e^3\Sigma_u^+)}$, and H(n=2) for the same collision systems. One more thing we can notice in fig. (\ref{fig:Q} (c,d)) that the width of individual peaks are more broader for $\mathrm{H_2^+}$ projectile. It is because more than one vibrational levels were populated in each electronic states.  At 0.5 a.u. relative collision velocity, our Q-value measurement resolution was $\mathrm{\delta Q \sim\ 2.7\ eV}$ which is much grater than vibrational level spacing, so they are unresolved. 

 $\mathrm{H_2^+ (X^2\Sigma_g^+) + H_2(X^1\Sigma_g^+) \rightarrow H_2(X^1\Sigma_g^+) + H_2^+ (X^2\Sigma_g^+)}$ is the symmetrical resonant charge transfer channel. $\mathrm{H^+ + H_2(X^1\Sigma_g^+) \rightarrow H(n=1) + H_2^+ (X^2\Sigma_g^+)}$ channel is near resonant. These two channels are of our interest in this study. We shall refer to these channels as the ground state channel, and the others as excited state channel.  To evaluate the branching ratio of these channels, we performed two peak fitting on the Q-value spectrum. The contribution of two peaks are shown in red dotted curve in fig. (\ref{fig:Q} (c,d)). From area under each peak, we deduced relative population of the ground state electron capture. The results are shown in fig. (\ref{fig:Q}(e)) and fig. (\ref{fig:Q}(f)). CTMC simulation reproduced this ratio for $\mathrm{H^+}$ projectile. It is clear from our result that in previously reported absolute differential cross section measurement by Gao $et\ al.$\cite{PhysRevA.44.5599}, the observed structure was 97$\%$ of the ground state channel. For $\mathrm{H_2^+}$, the relative population of the symmetrical resonant
 among all non-dissociative channel is more than $\mathrm{85 \%}$ in our projectile energy range. 

\subsection{Interference fringe pattern} 
\begin{figure}[h]
    \centering
    \includegraphics[width=\linewidth]{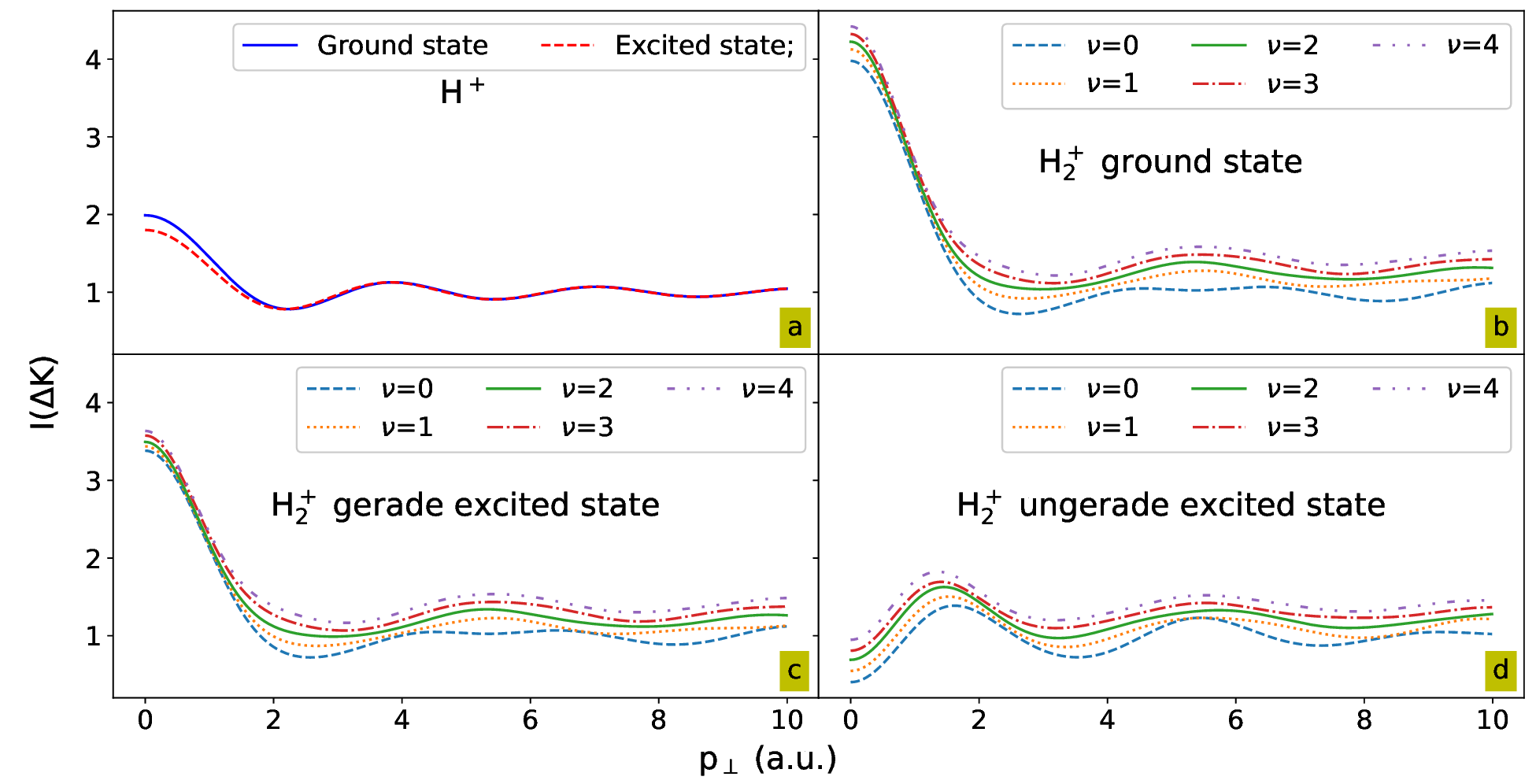}
    \caption{Orientation averaged interference patterns for 7.5 keV/u projectiles. Panel (a) presents patterns for $\mathrm{H^+ + H_2}$ collision. Panel (b) presents patterns for $\mathrm{H_2^+ + H_2}$ collision ground state channel. Panel (c), (d) presents patterns for gerade state and ungerade state transition, excited state channel (see table \ref{tab:q}). In Panel b,c,d results for first five vibrational levels are plotted with different offset for better visibility.}
    \label{fig:interference}
\end{figure}

The de-Broglie wavelength is three order of magnitude less than the internuclear separation (for 7.5 keV $\mathrm{H^+}$, $\mathrm{\lambda_{dB} }$ = 0.3 pm), As a result, each nuclei are acting as independent, coherent scattering centers \cite{book3,PhysRev.117.756,PhysRev.150.30}. In $\mathrm{H^+ + H_2}$ collision system, two single center scattering amplitude (Eqn. \ref{eqn:TH+H2}), and in $\mathrm{H_2^+ + H_2}$ collision system, four single center scattering amplitude (Eqn. \ref{eqn:TH2++H2}), will interfere.  We measured molecular orientation averaged differential cross sections in non-dissociative channels, and the interference patterns will be given by Eqn. \ref{eqn:intf_H2+} and \ref{eqn:intf_H2}. The double slit separation of $\mathrm{H_2}$ and $\mathrm{H_2^+}$ are the respective internuclear separations during the collision. Internuclear separation will follow the probability distribution that is square moduli of the vibrational wave function. $\mathrm{H_2}$ will always be in $\mathrm{X^1\Sigma_g^+;\nu = 0}$ state and $\mathrm{H_2^+}$ may be in any of $\mathrm{X^2\Sigma_g^+; \nu}$ states.  To calculate the interference fringe patterns using Eqn. \ref{eqn:intf_H2+}, we used vibrational level images reported by L.PH Schmidit $et\ al.$ \cite{PhysRevLett.108.073202}. The results are shown in fig. (\ref{fig:interference}). 

In fig. (\ref{fig:interference}) it is evident that for a given channel, the fringe structure changes negligibly with vibrational levels of $\mathrm{H_2^+}$. Thus we took only the fringe structure of the dominating vibrational level, $\nu$=2, to analyze the measured differential cross sections.We can notice the effect of constructive and destructive interference for transition to gerade and ungerade states in fig. (\ref{fig:interference} (c) and (d)).

\begin{table}[h]
    \centering
    \begin{minipage}[b]{0.48\textwidth}
        \centering
        \begin{tabular}{|c|c|}
        \hline
        \multicolumn{2}{|c|}{$\mathrm{H^+ + H_2}$}\\
        \hline
           Populated state  & Q value (eV) \\
           \hline
            n=1 & -2.4 \\
            \hline
            n=2 & -12.6\\
            \hline
        \end{tabular}
    \end{minipage}
    \hfill
    \begin{minipage}[b]{0.48\textwidth}
    \centering
    \begin{tabular}{|c|c|}
    \hline
    \multicolumn{2}{|c|}{$\mathrm{H_2^+ + H_2}$}\\
    \hline
           Populated state  & Q value (eV) \\
      \hline
     $\mathrm{X^1\Sigma_g^+, \nu=6}$ & -2.7\\
     \hline
       $\mathrm{C^1\Pi_u, \nu = 2}$ & -12.8 \\
      \hline
      $\mathrm{EF^1\Sigma_g^+, \nu = 6}$ & -12.8\\
      \hline
     $\mathrm{c^3\Pi_u, \nu = 2}$ & -12.3 \\
     \hline
     $\mathrm{a^3\Sigma_g^+, \nu=2}$ & -12.7 \\
     \hline
     $\mathrm{e^3\Sigma_u^+, \nu=2}$ & -13.7\\
     \hline
     \end{tabular}
    \end{minipage}
    
    \caption{Possible states populated after electron capture and corresponding Q-values. For electron capture by $\mathrm{H_2^+}$, most probable vibrational levels allowed by Frank Condon principle \cite{WUNDERLICH2011152} are quoted.}
    \label{tab:q}
\end{table}

\subsection{Differential cross section} 
\begin{figure}[H]
    \centering
    \includegraphics[width=0.99\textwidth]{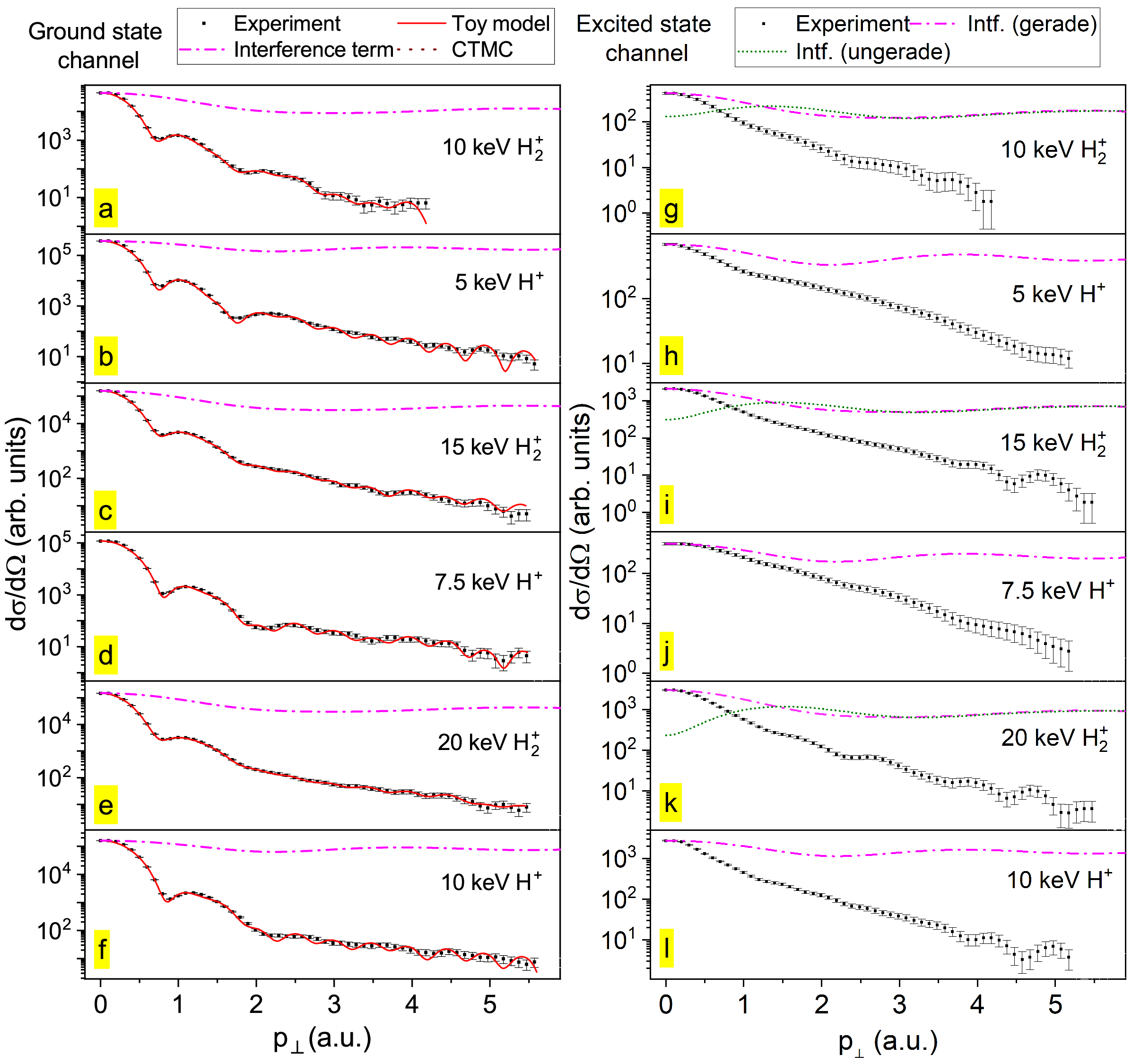}
    \caption{Channel resolved differential cross sections. Panel (a,b), (c,d), (e,f) presents results for ground state electron capture by equal velocity $\mathrm{H_2^+}$ and $\mathrm{H^+}$ ions. Panel (g-l) presents results for excited state electron capture. Black square represents experimental data, red solid curve represents toy model fitting results, wine dotted curve represents CTMC results. Magenta dash-dot curve and olive short-dot curves represents calculated orientation averaged interference fringes for gerade and ungerade state transitions. Error-bars are statistical in nature ($\mathrm{\sqrt{n}}$).}
    \label{fig:dndo}
\end{figure}

In fig. (\ref{fig:dndo}), the measured state selective differential cross sections for two resolved channels are plotted.  Our measurements lie in small scattering angle regime  where Fraunhofer like diffraction effect is expected to be seen \cite{PhysRevLett.87.123201,M_van_der_Poel_2002}.  The observed structure in differential cross section is modulation of a single slit diffraction pattern and an interference pattern, described by Eqn. \ref{eqn:dndo_H+} and \ref{eqn:dndo_H2+}. In fig. (\ref{fig:dndo}), we can notice that the molecular-orientation-averaged interference fringe visibility is much weaker compared to the main structures of the measured differential cross sections. This lead us to conclude that the interference fringe hardly modulate the single center diffraction pattern  and thus the observed structures are dominantly the single center diffraction patterns. Electron transfer to different electronic states diffract the matterwave in different ways. In this sense, the diffraction pattern in the ground state channel is unique. The  pattern in $\mathrm{H_2^+ + H_2}$ excited state channel is  addition of five different type diffraction patterns (see table \ref{tab:q}). If the final state is a gerade state, the interference pattern is constructive (magenta dash dot curve in fig. (\ref{fig:dndo} (g), (i), (k))), while for ungerade state transition, the interference pattern is destructive (gray dash-dot-dot curve in fig. (\ref{fig:dndo} (g), (i), (k))). 

Since the main objective of this study is to investigate resonant and near resonant charge exchange, henceforth we are discussing only the ground state channel. In this channel, $\mathrm{\Delta m =0}$, for both $\mathrm{H^+}$ and $\mathrm{H_2^+}$ projectiles, as electron transfer take place $\mathrm{H^+ : H_2(^1\Sigma_g^+) \rightarrow H(1s)}$ and $\mathrm{H_2^+ : H_2(^1\Sigma_g^+) \rightarrow H_2(^1\Sigma_g^+)}$. Assuming the two identical scattering centers in $\mathrm{H_2}$ and $\mathrm{H_2^+}$ act like atomic scatterers \cite{book3,PhysRev.117.756,PhysRev.150.30}, T-matrices $\mathrm{T_0}$ and $\mathrm{T_{00}}$, can be modeled by Eqn. \ref{eqn:fth}. The oscillatory nature in fig. (\ref{eqn:dndo} a-f) is emanating from oscillatory nature of Bessel function $\mathrm{J_0}$. In some earlier studies, this type oscillatory structure has also been observed in small angle, charge exchange scattering.  For example,  Gao $et\ al.$ \cite{PhysRevA.44.5599} observed identical structure in 5 keV $\mathrm{H^+ + H_2}$, absolute charge exchange cross section measurement. Though that study was not state selective, using our Q-value spectrum we can confirm that the ground state transfer probability was 97 $\%$,  thus the oscillatory structure of this channel was dominant. In other state selective electron capture studies, in $\mathrm{He^+ + H_2}$ \cite{PhysRevA.85.042704}, $\mathrm{Ar^{8+} + H_2}$ \cite{PhysRevA.109.032819}, S-shell electron capture also exhibited oscillatory differential cross section, that came from same diffraction effect. In $\mathrm{H_2^+ + H_2}$ symmetrical resonant transfer, previously observed distinct oscillatory nature at large scattering angle \cite{PhysRev.132.2078} is subdued by this diffraction effect in small angular range. McCarrol $et\ al.$ \cite{R_McCarroll_1970} examined this effect theoretically in impact parameter treatment on $\mathrm{H^+ + H}$ collision. 

A noticeable feature of the ground state channel diffraction patterns is that for equal velocity $\mathrm{H^+}$ and $\mathrm{H_2^+}$ projectiles, the fringe widths are equal in our measurement resolution.  This suggest that the individual scattering centers dominantly diffract the matterwave from an annular region. Notably, at large scattering measurements, electron capture by $\mathrm{H^+}$ from H and $\mathrm{H_2}$ targets also exhibited peaks around equal collision energies \cite{PhysRev.125.567,PhysRev.118.1552}. This suggests at close impact, something similar happen to identical scattering centers in $\mathrm{H_2}$ and isolated H.  More details can be found in the impact parameter dependence.

\subsection{Impact parameter dependence}
\begin{figure}[h]
    \centering
    \includegraphics[width=\linewidth]{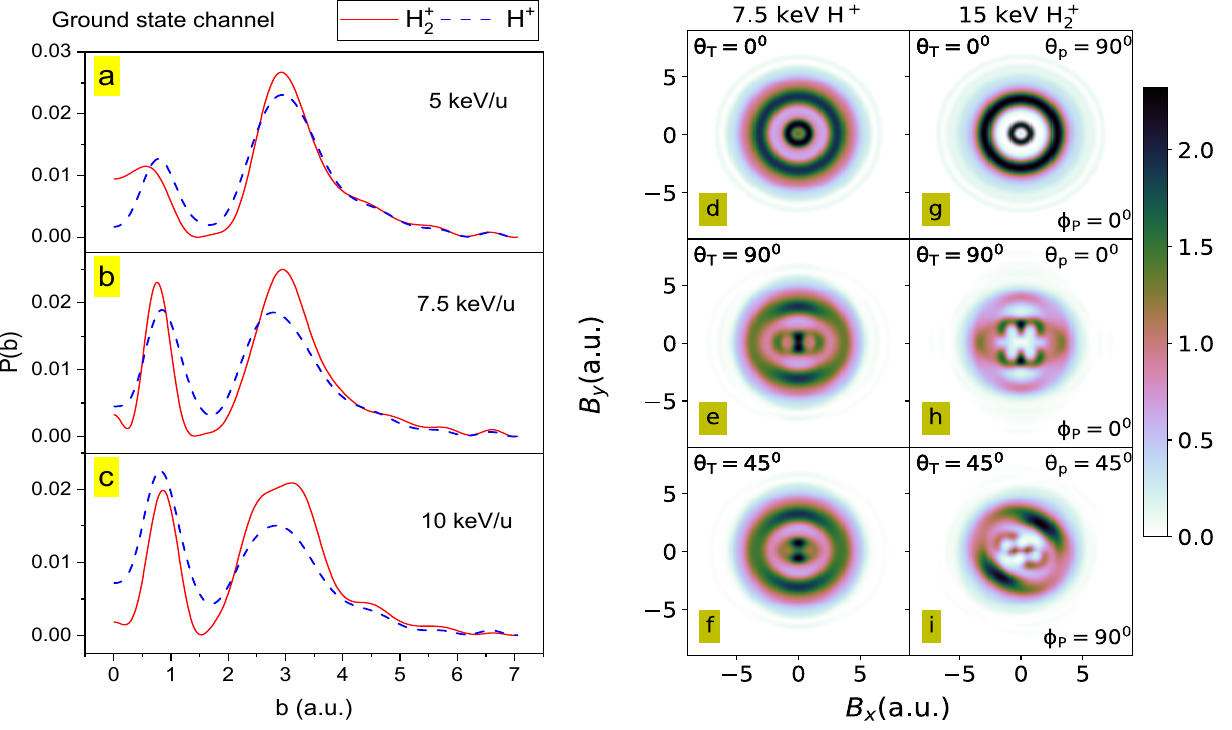}
    \caption{Impact parameter dependent probability of electron capture to ground state. Panel (a, b, c) presents results for individual scattering centers in equal velocity $\mathrm{H_2^+ + H_2}$ and $\mathrm{H^+ + H_2}$ collisions. Panel (d,e,f) presents reconstructed 2-dimensional double slits at different orientations for 7.5 keV $\mathrm{ H^+ + H_2}$ ground state channel. Panel (g,h,i) presents the same for 15 keV $\mathrm{ H_2^+ + H_2}$.}
    \label{fig:pb}
\end{figure}
We derived impact parameter dependence using our toy model in Fraunhofer diffraction analogy, for ground state channel. A complex circular aperture can be associated with each scattering centers. Profiles of those apertures are depicted in fig. (\ref{fig:pb} (a-c)).  The probability distribution P(b) exhibited a main peak around b = 2.8 a.u. and a secondary peak around b = 0.8 a.u..   Using this result, constructed double slit for $\mathrm{H^+ + H_2}$ and quadrupole slit for $\mathrm{H_2^+ + H_2}$ using Eqns.\ref{eqn:ab2d h + h2}, \ref{eqn:ab2d h2+ + h2}  are depicted in fig. (\ref{fig:pb} (d-i)). The qualitative structures of P(b) associated with individual scattering center P(b) are near identical for both the projectiles . This suggest that if we model the individual scattering centers of $\mathrm{H_2^+}$  with two independent identical atomic centers, electron transfer dynamics are similar with isolated $\mathrm{H^+}$ when colliding with $\mathrm{H_2}$.

Dynamics of $\mathrm{H^+ + H_2}$ can be understood from previously reported close coupling calculations \cite{PhysRevA.40.1302,PhysRevA.32.802}. Two adiabatic potential energy curves corresponds to entry channel $\mathrm{H^+ + H_2 : X^1\Sigma_g}$ and exit channel $\mathrm{H(1s) + H_2^+ : 1s\sigma_g}$,  do not any avoided crossing \cite{PhysRevA.32.802,PhysRevA.44.5599}, thus the transition may be classified as Demkov type \cite{demkov1964charge}. Calculated radial coupling by Kimura \cite{PhysRevA.32.802} suggest that Demkov-type transition take place between the incident channel  and exit channel  in a narrow region around b = 4.5 a.u. and the interference effect between two pathways take place for trajectories bellow b = 4.5 a.u.. The occurrence of same effect was mentioned by Gao $et\ al.$\cite{PhysRevA.44.5599}. 

About $\mathrm{H_2^+ + H_2}$ resonant channel, we may think that like other atomic collisions, periodic transition between symmetric and antisymmetric wavefunctions results the oscillatory nature \cite{PhysRev.131.229}. To closely examine the nature of these wave functions, derivation of potential energy curves  are required considering individual scattering centers as effective atoms.  Comparing impact parameter studies by Shingal and Lin \cite{R_Shingal_1989}, collision of $\mathrm{H^+}$ with H and $\mathrm{H_2}$ also exhibited a common peak position in P(b).  Similarity in electronic structures of H(1s) and $\mathrm{H_2(X^1\Sigma_g^+)}$ around one nuclei may be responsible for this.  

\section{Conclusion}
We measured state selective small angle differential cross sections in equal velocity $\mathrm{H^+ + H_2}$ and $\mathrm{H_2^+ + H_2}$ collisions. The ground state electron capture by $\mathrm{H_2^+}$ is resonant channel and by $\mathrm{H^+}$ is a near resonant channel. We analyzed the results by exploiting  multi scattering center concept. We found that in non-dissociative state electron capture, the orientation averaged multi center interference pattern very weakly modulate the single center diffraction pattern.  We interpreted our measured results using impact parameter dependence derived by a toy model in Fraunhofer diffraction analogy. We found that the impact parameter dependent probability of electron capture individual scattering centers in equal velocity $\mathrm{H^+ + H_2}$ and $\mathrm{H_2^+ + H_2}$ collisions are near identical. This implies that the dynamics of molecular resonant electron capture by $\mathrm{H_2^+}$ has some similarities with constituent atomic ion $\mathrm{H^+}$.

\bibliographystyle{ieeetr}
\bibliography{apssamp}

\end{document}